\documentclass[10pt,conference]{IEEEtran}
\usepackage{amsfonts}
\usepackage{dsfont}
\usepackage{setspace}

\usepackage{setspace}
\usepackage{amsmath}
\usepackage{amssymb}
\usepackage[dvips]{graphicx}
\usepackage{epsfig}
\usepackage{dsfont}

\usepackage[ps2pdf,
bookmarks=false,
bookmarksnumbered=false, 
bookmarksopen=false, 
colorlinks=false]{}

\newcommand{\E}{\mathbb{E}}

\begin{document}

\title{Intercell Interference-Aware Scheduling for Delay Sensitive Applications in C-RAN}

\author{Yi Li, M. Cenk Gursoy and Senem Velipasalar
\\Department of Electrical Engineering and Computer Science,
Syracuse University, Syracuse, NY 13244
\\Email: yli33@syr.edu, mcgursoy@syr.edu, svelipas@syr.edu}

\maketitle

\begin{abstract}\let\thefootnote\relax\footnote{This work was supported in part by National Science Foundation grants CCF-1618615, ECCS-1443994, CNS-1443966, CNS-1302559, and CNS-1206291.}
Cloud radio access network (C-RAN) architecture is a new mobile network architecture that enables cooperative baseband processing and information sharing among multiple cells and achieves high adaptability to nonuniform traffic by centralizing the baseband processing resources in a virtualized baseband unit (BBU) pool. In this work, we formulate the utility of each user using a convex delay cost function, and design a two-step scheduling algorithm with good delay performance for the C-RAN architecture. In the first step, all users in multiple cells are grouped into small user groups, according to their interference levels and estimated utilities. In the second step, channels are matched to the user groups to maximize the system utility. The performance of our algorithm is further studied via simulations, and the advantages of C-RAN architecture is verified.
\end{abstract}

\section{Introduction}
Recently, the traffic load of wireless cellular networks has grown dramatically due to increasing number of smart mobile devices. In order to satisfy the growing demands and provide the required quality of service (QoS) guarantees and high reliability in next-generation 5G wireless systems, several advanced techniques have been proposed, and cloud radio access network (C-RAN) is one novel mobile network architecture that improves the performance of cellular networks. By centralizing the baseband processing resources of multiple cells in a virtualized baseband unit (BBU) pool, C-RAN can achieve cooperative processing among different cells and utilize the BBUs more efficiently \cite{CRAN_01} \cite{mobile2011c}. As shown in Figure \ref{fig:system}, remote radio heads (RRHs) and BBU are separated geographically and connected via optical fibers in the C-RAN architecture. BBU pool is shared between cells as a virtualized cluster. Compared with the conventional architectures in which BBUs of different cells are not shared, C-RAN can achieve information exchange and cooperative processing between cells more easily with low latency, and it has high adaptability to nonuniform traffic. A comprehensive survey on C-RAN and its implementation is provided in \cite{CRAN_survey}.

For most orthogonal frequency division multiple access (OFDMA)-based cellular networks, intercell interference (ICI) is a significant interference source because of the frequency reuse among multiple neighbouring cells. Many advanced methods have been studied to control ICI. For instance, the soft frequency reuse (SFR) scheme is proposed in \cite{SFR_01} and \cite{SFR_02}, in which cell edge users transmit with high power in non-overlapping cell edge bands allocated to adjacent cells, and center users use the cell center bands with limited transmission power. The authors in \cite{xiang2007inter} further compared the performance of SFR with partial frequency reuse scheme. In these conventional ICI control schemes, cooperation between neighbouring cells are not considered, which limits their performances. In C-RAN, cooperative processing among the cells sharing the same BBU pool becomes easier and more efficient, which helps to improve ICI control. In \cite{ICI_HetNets}, a resource allocation and RRH association algorithm was proposed for ICI coordination in a long term evolution (LTE) heterogeneous network setting with C-RAN architecture. However, optimization over multiple cells greatly increases the complexity, which causes problems in delay sensitive applications. In this work, we propose, for C-RAN, an ICI-aware scheduling algorithm that controls the ICI with relatively low complexity.

In addition, packet delay is an important performance criterion for delay sensitive applications such as live video streaming and online gaming. In most of the related studies considering ICI control, the objectives are interference minimization, SINR maximization and throughput maximization, and hence delay minimization is not addressed. In this work, our scheduling algorithm performs user grouping and resource allocation with the goal of minimizing the delay violation probability. The utility formulation used in this paper has also been employed in our previous work \cite{D2D_scheduling}.

The main contributions in this paper are listed as follows:
\begin{enumerate}
  \item We propose a two-step ICI-aware scheduling algorithm for C-RAN that minimizes the delay violation probability of the system.
  \item We design a novel user grouping algorithm for the user grouping step, which controls the interference among the users in the same group.
  \item We formulate the channel assignment problem in the second step as a maximum-weight matching problem, which can be solved using standard algorithms in graph theory.
  \item We verify the performance of our algorithm via simulations, and compare our algorithm with a conventional soft frequency reuse (SFR) algorithm. Also, the influence of the system parameters is investigated with the help of numerical results.
\end{enumerate}

\section{System Model and Preliminaries}\label{Sec:Preliminary}
In this section, we introduce our system model in the first subsection, and subsequently describe the utility formulation used in this work in the second subsection.

\subsection{System Model} \label{sec:model}
\begin{figure}
\begin{center}
\hspace{-1cm}
\includegraphics[width=0.5\textwidth]{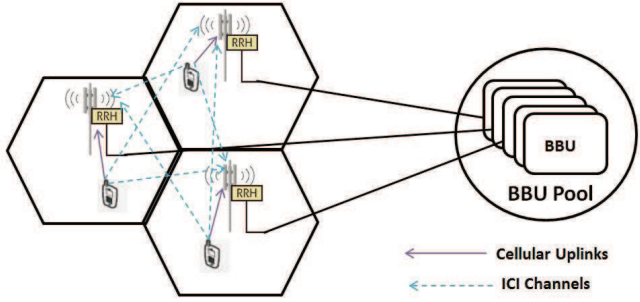}
\caption{System model of C-RAN with ICI}\label{fig:system}
\end{center}
\end{figure}

In this work, the uplink transmission in a C-RAN within an OFDMA setting is considered as shown in Fig. \ref{fig:system}. There are $N_c$ cells in this network, and each cell is served by a base station with one RRH. RRHs are connected to a centralized BBU pool with multiple BBUs working cooperatively. All cells reuse $N_{ch}$ frequency bands/channels, and each channel has a bandwidth of $B$. The total number of mobile users in this network is fixed at $N_u$, and users are assumed to be associated with their nearest RRHs. Each user is equipped with a buffer storing the arriving packets before sending them through the wireless uplink channels, and the size of each packet is assumed to be $I_p$ bits. All buffers are assumed to operate in a first-in first-out (FIFO) manner. The system is assumed to operate under delay constraints, and target delay of packets sent by the $i^{\text{th}}$ user is denoted by $D_i$ (time frames). Block fading is assumed in this work, in which the fading coefficients stay constant within one time frame with a duration of $T$, and change across frames. Also it is assumed that the distributions of the fading coefficients are identical in different channels.

At the beginning of each time frame, BBU pool allocates channel resources to the users using a scheduling algorithm. It is assumed that users keep silent until they get channel resources from the BBU pool, and the channel resources are returned back at the end of each time frame. There are $4$ assumptions for the channel assignment:
\begin{enumerate}
  \item The number of users is much greater than the number of available channels, $N_u\gg N_{ch}$. In such a case, each user transmits using one channel at most.
  \item Only the users that can satisfy the pair-wise interference constraints given in (\ref{eq:interference_constraint}) can reuse the same channel resource.
  \item Users associated with the same RRH cannot reuse the same channel resource.
  \item The BBU pool is assumed to have perfect channel side information (CSI), and it is also assumed to keep track of the buffer status (including the queue length and packet delay information) of each user.
\end{enumerate}
The first assumption addresses a heavy load scenario, in which all channels are reused by multiple users and ICI becomes a significant problem. In such a case, the assumption that each user transmits using one channel at most helps to reduce ICI caused by excessive frequency reuse. The second assumption limits the interference, and the third assumption guarantees that all interference comes from neighbouring cells. The last assumption guarantees that the BBU pool has enough information to conduct our scheduling algorithm. CSI is estimated at RRHs and sent to the BBU pool via optical fiber links. Information of the arrival rates at all users is also sent to the BBU pool via special feedback channels\footnote{We assume ideal feedback without delay and error.}, and the BBU pool can track the queue status at each user.

Define $\bold{\Psi}_{j}(t)$ as the set of users that use the $j^{\text{th}}$ channel in the $t^{\text{th}}$ time frame, and $\xi_{i,j}(t)$ as the indicator function that indicates whether the $j^{\text{th}}$ channel is assigned to the $i^{\text{th}}$ user in the $t^{\text{th}}$ time frame. In other words, $\xi_{i,j}(t)=1$ if $i\in\bold{\Psi}_{j}(t)$, otherwise $\xi_{i,j}(t)=0$. According to our first channel assignment assumption, we have $\sum_{j=1}^{N_{ch}}\xi_{i,j}(t)\leq 1$. Then for the $t^{\text{th}}$ time frame, the received signal corresponding to user $i$ at its associated base station can be expressed as
\begin{small}
\begin{align}
y_i=h^j_{i}x_i+\sum_{k\in\bold{\Psi}_{j}(t),k\neq i}h^j_{k,i}x_k+n^j_i
\end{align}
\end{small}
if $\xi_{i,j}(t)=1$. Above, $x_i$ represents the transmitted signal of user $i$, $h^j_{i}$ denotes the fading coefficient of the channel between user $i$ and its corresponding RRH, $h^j_{k,i}$ denotes the fading coefficient of the interference channel between user $k$ and the RRH associated with user $i$, and $n^j_i$ is the background noise at the base station associated with user $i$ which is assumed to follow an independent complex Gaussian distribution with zero mean and variance $\sigma^2$, i.e., $n^j_i \sim \mathcal{CN}(0,\sigma^2)$. The transmission rate of user $i$ in the $t^{\text{th}}$ time frame is given by
\begin{small}
\begin{align}\label{eq:inst_rate}
r_i(t)=TB\log_2\left(1+\frac{P_i z^j_i}{B\sigma^2+\sum_{k\in\bold{\Psi}_{j}(t),k\neq i}P_k z^j_{k,i}} \right)\:\text{bits/frame}
\end{align}
\end{small}
where $j$ is the index of the channel that is assigned to user $i$, $P_i$ represents the transmission power of user $i$, $T$ is the duration of each time frame, $B$ is the bandwidth of each channel, $z^j_i=|h^j_{i}|^2$, and $z^j_{k,i}=|h^j_{k,i}|^2$.

\subsection{Convex Delay Cost and Utility}
In the convex delay cost approach, the cost function of a packet is formulated as an increasing convex function of its delay \cite{van_dynamic}. The high performance of this approach was shown in \cite{van_due} for a single cell model without any interference. In our previous work \cite{D2D_scheduling}, we designed a scheduling algorithm using the convex cost function provided in \cite{van_due} for a D2D communication setting, and verified via simulations that this approach has very good delay performance. Here, we define the cost of the $j^{\text{th}}$ packet in the buffer at user $i$ as
\begin{small}
\begin{align}
C_{j,i}=\frac{d_{j,i}}{D_i},
\end{align}
\end{small}
where $d_{j,i}$ is the current delay of this packet, and $D_i$ is the target delay of user $i$. At user $i$, the number of packets that can be transmitted in the current time frame is
\begin{small}
\begin{align}\label{eq:mu}
\mu_i=\min\left\{l_i, \lfloor r_i/I_p\rfloor\right\},
\end{align}
\end{small}
where $l_i$ is the number of packets waiting in the buffer at user $i$, $I_p$ is the size of each packet, and $\lfloor\cdot\rfloor$ represents the floor function. The utility of user $i$ is defined as
\begin{small}
\begin{align}\label{eq:utility_user}
U_i=\sum_{j=1}^{\mu_i}C_{j,i},
\end{align}
\end{small}
and the utility of the system is defined as
\begin{small}
\begin{align}\label{eq:utility}
U=\sum_{i=1}^{N_u} U_i =\sum_{i=1}^{N_u}\sum_{j=1}^{\mu_i}C_{j,i}.
\end{align}
\end{small}
The utility given in (\ref{eq:utility}) represents the total cost of the packets that can be transmitted to the base station in the current time frame. At the beginning of each time frame, the BBU pool runs a scheduling algorithm for channel assignment to maximize the utility. In the next section, a detailed discussion on our scheduling algorithm is provided.

\section{ICI-Aware Scheduling Algorithm for C-RAN}
In this section, we introduce our scheduling algorithm. In each time frame, our scheduling algorithm assigns channels to the users in a way that maximizes the utility given in (\ref{eq:utility}). Since we consider a C-RAN architecture, the BBU pool has the knowledge of all fading distributions and cost functions of each packets, and it can allocate channel resources to all users in different cells together. Our scheduling algorithm can be divided into two steps, namely the user grouping step and channel matching step. In the first step, we divide all users into small groups such that the users in the same group reuse the same channel. In the second step, we match the channels to the user groups to maximize the utility.
\subsection{User Grouping}
In the first step of our algorithm, we divide all users into small groups, and each group will be assigned a channel resource in the next step. Before channel assignment, we cannot compute the instantaneous transmission rates because the sets $\bold{\Psi}_1$, $\bold{\Psi}_2$, $\cdots$, $\bold{\Psi}_{N_{ch}}$ have not been determined yet. Therefore, we use a rate estimator
\begin{small}
\begin{align}\label{eq:r_estimator}
\hat{r}_i=\frac{1}{m}\sum_{\tau=t-m}^{t-1}r_i(\tau)
\end{align}
\end{small}
instead. This rate estimator is essentially the average rate over the most recent $m$ time frames. Plugging (\ref{eq:r_estimator}) into (\ref{eq:mu}) and (\ref{eq:utility_user}), we obtain the utility estimator of user $i$ as
\begin{small}
\begin{align}\label{eq:utility_estimator}
\hat{U}_i=\sum_{j=1}^{\hat{\mu}_i}C_{j,i}=\sum_{j=1}^{\min\{l_i, \lfloor \hat{r}_i/I_p\rfloor\}}C_{j,i}.
\end{align}
\end{small}

In order to control ICI, we assume that any two users ($i_1$ and $i_2$) reusing the same channel resource have to satisfy the pairwise interference/SINR constraints given by
\begin{small}
\begin{align}\label{eq:interference_constraint}
\begin{cases}
\E\left\{\frac{P_{i_1}z_{i_1}}{B\sigma^2+P_{i_2}z_{i_2,i_1}}\right\}\geq\gamma\E\left\{\frac{P_{i_1}z_{i_1}}{B\sigma^2}\right\}\\
\E\left\{\frac{P_{i_2}z_{i_2}}{B\sigma^2+P_{i_1}z_{i_1,i_2}}\right\}\geq\gamma\E\left\{\frac{P_{i_2}z_{i_2}}{B\sigma^2}\right\}
\end{cases},
\end{align}
\end{small}
where the parameter $\gamma$ is between $0$ and $1$. Since the distributions of the fading coefficients are identical in different channels, the expected values of the SINRs and SNRs in (\ref{eq:interference_constraint}) do not depend on the channel assignment result. The details of our user grouping algorithm is given in Table \ref{Algorithm1}, and we denote the number of the output user groups as $N_g$.
\begin{table}
\footnotesize
\caption{\label{Algorithm1} User Grouping Algorithm}
\begin{tabular}{p{8.4cm}}
\hline
\hline
\textbf{Input}: $\gamma$, transmission power and utility estimator of each user, the fading coefficients.\\
\textbf{Output}: User groups $GP_1$, $GP_2$, $\cdots$, $GP_{N_g}$.\\
\hline
\hline
Collect the utility estimators $\hat{U}_i$ into a vector $\bold{V}=[\hat{U}_1,\hat{U}_2,\cdots,\hat{U}_{N_u}]$.\\
Set $k=1$\\
\textbf{While} $\max(\bold{V})\geq 0$\\
\hspace{0.5cm} Set $\bold{V}^*=\bold{V}$ and $GP_k=\emptyset$\\
\hspace{0.5cm} \textbf{While} $\max(\bold{V}^*)\geq 0$\\
\hspace{1cm} $i=\arg\max(\bold{V}^*)$\\
\hspace{1cm} Add user $i$ into $GP_k$.\\
\hspace{1cm} Set $\bold{V}(i)=-1$ and $\bold{V}^*(i)=-1$.\\
\hspace{1cm} \textbf{For} $j$ from $1$ to $N_u$\\
\hspace{1.5cm} Set $\bold{V}^*(j)=-1$ if user $i$ and $j$ cannot satisfy the\\
\hspace{1.5cm} interference constraints given in (\ref{eq:interference_constraint}) or they are associated\\
\hspace{1.5cm} to the same RRH.\\
\hspace{1cm} \textbf{End}\\
\hspace{0.5cm} \textbf{End}\\
\hspace{0.5cm} $k=k+1$\\
\textbf{End}\\
\hline\hline
\end{tabular}
\end{table}

At the beginning, we set group $GP_k$ as an empty set. Each time, we select the user with the maximum utility estimator and include it into $GP_k$. After adding a user into a group, we kick out the users that cannot reuse the same channel resource with this selected user by setting $\bold{V}^*(j)=-1$, which can be processed in parallel at the BBU pool. Our grouping algorithm aims to collect the users with high utility estimators together, which helps to serve these users with less channel resources.

Note that the number of groups $N_g$ might be smaller than the number of channels $N_{ch}$. In such cases, some of the channels cannot be assigned to users, and we need to break those groups with large sizes into several small groups so that $N_g=N_{ch}$. To divide a big group into two small groups, we select half of the users with smaller utility estimator values within the large group, and let them form a new small group.
\subsection{Channel Matching}
In the second step, we assign channels to the user groups via the maximum-weight matching approach. In this step, we find a matching between user groups and channels that maximizes the system utility given in (\ref{eq:utility}). Let us define $\eta_{i,j}$ as the indicator of the channel assignment result, i.e., $\eta_{i,j}=1$ if channel $j$ is assigned to $GP_i$, and $\eta_{i,j}=0$ if channel $j$ is not matched to $GP_i$. Then the matching problem can be formulated as
\begin{small}
\begin{align}
&\textbf{Maximize}_{\quad\eta_{i,j}}\hspace{1.2cm} U\notag\\
&\textbf{Subject to} \hspace{1.3cm} \eta_{i,j}\in \{0,1\}\notag\\
&\hspace{2.8cm} \sum_{j=1}^{N_{ch}}\eta_{i,j}\leq 1\notag\\
&\hspace{2.8cm} \sum_{i=1}^{N_g}\eta_{i,j}=1\notag .
\end{align}
\end{small}

In graph theory, the maximum-weight matching problem can be solved by the Hungarian algorithm (Kuhn-Munkres algorithm) \cite{kuhn}. To use the Hungarian algorithm, we have to first construct the utility matrix $\bold{U}$, in which each row corresponds to a user group and each column corresponds to a channel. The element of this matrix $U_{i,j}$ is the sum utility of the users in $GP_i$ if the $j^{\text{th}}$ channel is assigned to that group. The elements of the utility matrix can be computed in parallel at the BBU pool. After constructing the utility matrix, the Hungarian algorithm is applied, and channels are assigned to the users.

\subsection{Summary and Complexity Analysis}
In summary, we propose a two-step scheduling algorithm with good delay performance for a multi-cell C-RAN model. In the first step, we group the users to control the ICI and aim to collect the users with high utility estimator values into smaller number of groups. In the second step, we formulate the channel allocation problem as a maximum-weight matching problem, and assign the channel resources to the user groups using the Hungarian algorithm. Although our algorithm only considers an uplink scenario, it can also be easily adapted to a downlink scenario.

Since we consider a C-RAN model, our algorithm is performed considering users in multiple cells, and parallel processing can be performed in some parts of our algorithm at the BBU pool to reduce time consumption. Compared with conventional resource allocation algorithms, in which cooperative processing among multiple cells is not considered, our algorithm has a significant potential to achieve better performance.

Assume that the number of processers at BBU pool is $\Theta(N_c)$, then the time complexity of the user grouping step is $O(N_u^2/N_c)$. In the matching step, the time consumption for constructing the utility matrix is $O(N_gN_{ch}/N_c)$, and the time consumption of the Hungarian algorithm is $O(\max\{N_g,N_{ch}\}^3)$. To further accelerate this process, we can replace the Hungarian algorithm with some heuristic algorithms with time complexity of $O(\min\{N_g,N_{ch}\})$. As an example, in each iteration, we can select the maximum element in the utility matrix, and match its corresponding group and channel together. The overall time consumption of this algorithm depends on the relationship among $N_u$, $N_c$, $N_g$ and $N_{ch}$.

\section{Numerical Results}
In this section, we further study the performance of our algorithm and the influence of parameters via simulations. In our simulations, we consider a C-RAN with $3$ adjacent cells, each with a radius of $2$. The coordinates of the RRHs of these three cells are $(-2,0)$, $(0,2)$ and $(2,0)$, respectively. In each cell, there are $5$ randomly placed users, and each one has the maximum transmission power $\frac{P_{max}}{B\sigma^2}=13$ dB. The number of available channels is $N_{ch}=5$. We assume Rayleigh fading with path loss $\E\{z\}=s^{-4}$, where $s$ represents the distance between the transmitter and the receiver. Each point on the curves is determined by taking the average over the results of $500$ systems with randomly placed users, and the performance result of each system is evaluated over $5\times 10^4$ time frames.

\begin{figure}
\begin{center}
\includegraphics[width=0.4\textwidth]{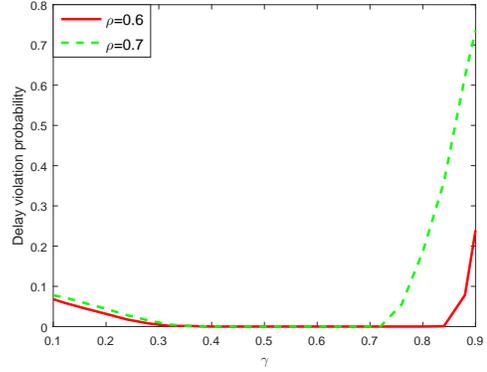}
\caption{Delay violation probability vs. interference control parameter $\gamma$}\label{fig:fig1}
\end{center}
\end{figure}

\begin{figure}
\begin{center}
\includegraphics[width=0.4\textwidth]{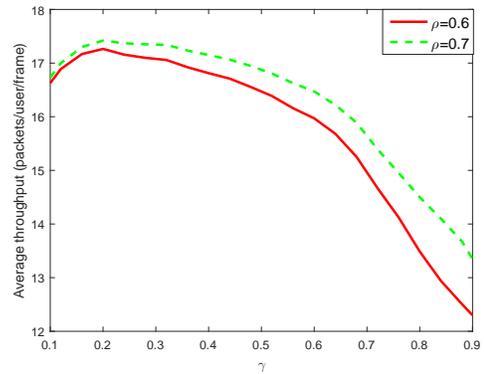}
\caption{Throughput vs. interference control parameter $\gamma$}\label{fig:fig2}
\end{center}
\end{figure}

In Figs. \ref{fig:fig1} and \ref{fig:fig2}, we study the influence of the interference control parameter $\gamma$, which is used in the pairwise interference constraints expressed in (\ref{eq:interference_constraint}). The arrival rate at user $i$ is set as $\lambda_i=\rho\E\{TB\log_2(1+P_iz_i/B\sigma^2)\}$, where the parameter $\rho$ is the arrival intensity. The target delay is $25$ time frames for all users, and all users transmit at their maximum power level. When $\gamma$ is small, the ICI is not well controlled and the average transmission rate is not maximized. As $\gamma$ increases, the system achieves lower delay violation probability and higher throughput due to better ICI management. However, when $\gamma$ is too large, the interference constraints become too strict, which leads to less frequency reuse. In such cases, the throughput becomes smaller and the delay violation probability increases.

\begin{figure}
\begin{center}
\includegraphics[width=0.4\textwidth]{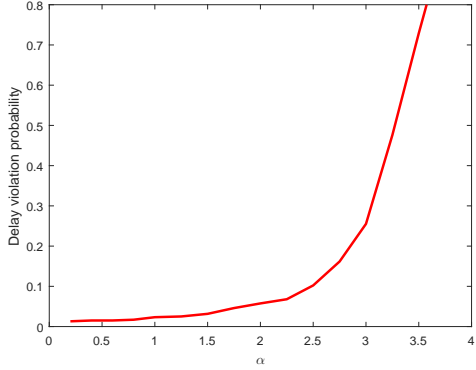}
\caption{Delay violation probability vs. power control parameter $\alpha$}\label{fig:fig3}
\end{center}
\end{figure}

\begin{figure}
\begin{center}
\includegraphics[width=0.4\textwidth]{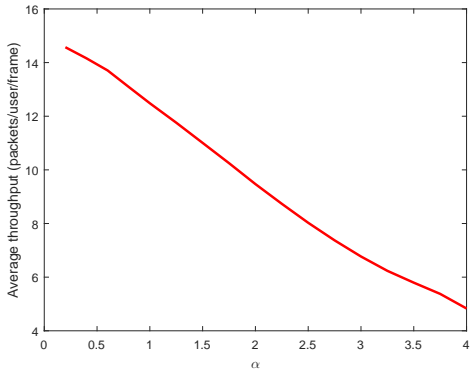}
\caption{Throughput vs. power control parameter $\alpha$}\label{fig:fig4}
\end{center}
\end{figure}
In Figs. \ref{fig:fig3} and \ref{fig:fig4}, we analyze the influence of power control on our algorithm. In several conventional ICI control algorithms such as SFR, cell center users transmit with small power to reduce the interference they cause to the cell edge users. We adopt this strategy and apply it in our algorithm. In these two figures, the transmission power of user $i$ is selected as $P_i=P_{max}(s_i/R_{cell})^\alpha$, where $s_i$ is the distance between the user and its corresponding RRH, and $R_{cell}$ is the radius of the cell. As $\alpha$ increases, cell center users are restricted to transmit with smaller power. Also, all arrival rates are set as $\lambda=1.5\E\{TB\log_2(1+P_{max}z_{edge}/B\sigma^2)\}$, where $\E\{TB\log_2(1+P_{max}z_{edge}/B\sigma^2)\}$ is the average transmission rate of a user at the edge of its associated cell. In Figs. \ref{fig:fig3} and \ref{fig:fig4}, we notice that as $\alpha$ increases, both delay and throughput performances become worse. Our algorithm control the interference in the user grouping step. Users that cannot satisfy the pairwise interference constraints are not allowed to reuse the same channel resource. Further decrease in the transmission power of the cell center users reduces their transmission rates, making it more difficult to stabilize the system.

\begin{table}
\begin{center}
\caption{Comparison between our algorithm and SFR}\label{table:table1}
\includegraphics[width=0.5\textwidth]{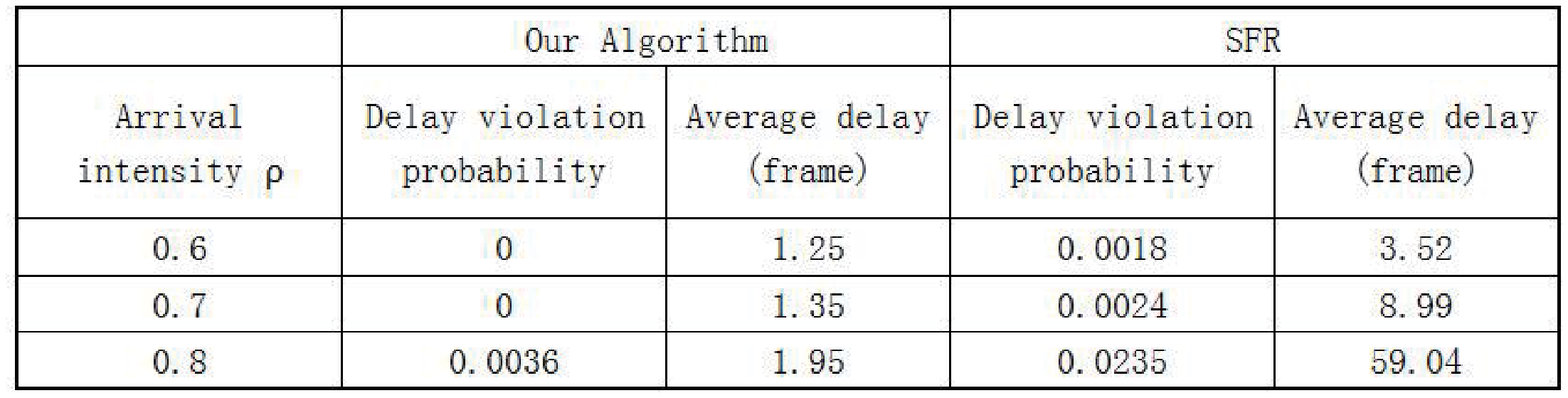}
\end{center}
\end{table}

Finally, we compare our algorithm with the conventional SFR scheme introduced in \cite{xiang2007inter}. The arrival rates are set in the same way as in Figs. \ref{fig:fig1} and \ref{fig:fig2}, and the target delay is $40$ for all users. In our algorithm, all users transmit with maximum power. In the SFR scheme, users transmit with full power in the edge bands and they use $70\%$ of their maximum power in the center bands. Channel assignment is conducted at the BBU of each cell individually to maximize the sum utility of the users in that cell. The results are provided in Table \ref{table:table1}. As the arrival intensity increases, the advantage of our algorithm becomes obvious in terms of the average delay. With the C-RAN architecture, cooperative processing over multiple cells enhances the delay performance significantly.

\section{Conclusion}
In this work, we have proposed an ICI-aware scheduling algorithm for the C-RAN architecture that minimizes the sum delay cost of the system. The procedure is divided into two steps, namely the user grouping step and the channel matching step. In the user grouping step, we have designed a grouping algorithm that partitions all users in the network into small groups by checking their pairwise interference levels. In order to serve those users with high utility values with less channel resources, our grouping algorithm aims to collect users with high utility estimator values into small number of groups. In the channel matching step, we have formulated the channel assignment problem as a maximum-weight matching problem, which can be solved using the Hungarian algorithm. In the second step, user groups are matched to the available channel resources with goal of maximizing the system utility. Finally, we have studied the impact of the interference threshold and power control parameter via simulations, and compared our algorithm to the conventional SFR scheme. With the advantages of cooperative processing and information sharing over multiple cells, it has been verified that our algorithm designed for C-RAN can achieve higher throughput and lower delay.

\bibliographystyle{ieeetr}
\bibliography{CRAN}

\begin{thebibliography}{10}

\bibitem{CRAN_01}
Y.~Lin, L.~Shao, Z.~Zhu, Q.~Wang, and R.~K. Sabhikhi, ``Wireless network cloud:
  Architecture and system requirements,'' {\em IBM Journal of Research and
  Development}, vol.~54, pp.~4:1--4:12, Jan. 2010.

\bibitem{mobile2011c}
C.~Mobile, ``{C-RAN}: {The} road towards green {RAN},'' {\em White Paper, ver},
  vol.~2, Oct. 2011.

\bibitem{CRAN_survey}
A.~Checko, H.~L. Christiansen, Y.~Yan, L.~Scolari, G.~Kardaras, M.~S. Berger,
  and L.~Dittmann, ``Cloud {RAN} for mobile networks - {A} technology
  overview,'' {\em IEEE Communications Surveys Tutorials}, vol.~17,
  pp.~405--426, Firstquarter 2015.

\bibitem{SFR_01}
A.~Simonsson, ``Frequency reuse and intercell interference co-ordination in
  {E-UTRA},'' in {\em IEEE 65th Vehicular Technology Conference (VTC) Spring},
  pp.~3091--3095, Apr. 2007.

\bibitem{SFR_02}
M.~C. Necker, ``Local interference coordination in cellular {OFDMA} networks,''
  in {\em IEEE 66th Vehicular Technology Conference (VTC) Fall},
  pp.~1741--1746, Sep. 2007.

\bibitem{xiang2007inter}
Y.~Xiang, J.~Luo, and C.~Hartmann, ``Inter-cell interference mitigation through
  flexible resource reuse in {OFDMA} based communication networks,'' in {\em
  European wireless}, vol.~2007, pp.~1--7, Apr. 2007.

\bibitem{ICI_HetNets}
S.~Deb, P.~Monogioudis, J.~Miernik, and J.~P. Seymour, ``Algorithms for
  enhanced inter-cell interference coordination ({eICIC}) in {LTE} {HetNets},''
  {\em {IEEE/ACM} Trans. Networking}, vol.~22, pp.~137--150, Feb. 2014.

\bibitem{D2D_scheduling}
Y.~Li, M.~C. Gursoy, and S.~Velipasalar, ``Scheduling in {D2D} underlaid
  cellular networks with deadline constraints,'' in {\em IEEE Vehicular
  Technology Conference (VTC) Fall}, Sep. 2016.

\bibitem{van_dynamic}
J.~A. Van~Mieghem, ``Dynamic scheduling with convex delay costs: {The}
  generalized $c\mu$ rule,'' {\em The Annals of Applied Probability},
  pp.~809--833, Aug. 1995.

\bibitem{van_due}
J.~A. Van~Mieghem, ``Due-date scheduling: {Asymptotic} optimality of
  generalized longest queue and generalized largest delay rules,'' {\em
  Operations Research}, vol.~51, pp.~113--122, Feb. 2003.

\bibitem{kuhn}
H.~W. Kuhn, ``The hungarian method for the assignment problem,'' {\em Naval
  Research Logistics Quarterly}, vol.~2, pp.~83--97, Mar. 1955.

\end{thebibliography}

\end{document}